# IMPACT OF THE OPTIMUM ROUTING AND LEAST OVERHEAD ROUTING APPROACHES ON MINIMUM HOP ROUTES AND CONNECTED DOMINATING SETS IN MOBILE AD HOC NETWORKS

Natarajan Meghanathan

Jackson State University, 1400 Lynch St, Jackson, MS, USA
natarajan.meghanathan@jsums.edu

## ABSTRACT

*Communication protocols for mobile ad hoc networks (MANETs) follow either an Optimum Routing Approach (ORA) or the Least Overhead Routing Approach (LORA): With ORA, protocols tend to determine and use the optimal communication structure at every time instant; whereas with LORA, a protocol tends to use a chosen communication structure as long as it exists. In this paper, we study the impact of the ORA and LORA strategies on minimum hop routes and minimum connected dominating sets (MCDS) in MANETs. Our primary hypothesis is that the LORA strategy could yield routes with a larger time-averaged hop count and MCDS node size when compared to the minimum hop count of routes and the node size of the MCDS determined using the ORA strategy. Our secondary hypothesis is that the impact of ORA vs. LORA also depends on how long the communication structure is being used. Our hypotheses are evaluated using extensive simulations under diverse conditions of network density, node mobility and mobility models such as the Random Waypoint model, City Section model and the Manhattan model. In the case of minimum hop routes, which exist for relatively a much longer time compared to the MCDS, the hop count of routes maintained according to LORA, even though not dramatically high, is appreciably larger (6-12%) than those maintained according to ORA; on the other hand, the number of nodes constituting a MCDS maintained according to LORA is only at most 6% larger than the node size of a MCDS maintained under the ORA strategy.*

## 1. INTRODUCTION

A mobile ad hoc network (MANET) is a dynamic distributed system of wireless nodes that move independently of each other. Routes in MANETs are often multi-hop in nature due to the limited transmission range of the battery-operated wireless nodes. MANET routing protocols are of two types [1][2]: proactive and reactive. Proactive routing protocols determine routes between every pair of nodes in the network, irrespective of their requirement. Reactive or on-demand routing protocols determine routes between any pair of nodes only if data needs to be transferred between the two nodes and no route is known between the two nodes. Proactive routing protocols always tend to maintain optimum routes between every source-destination (*s-d*) pair and this strategy is called the Optimum Routing Approach (ORA) [1][3]. In this pursuit, each node periodically exchanges its routing table and link state information with other nodes in the network, thus generating a significantly larger control overhead. On the other hand, reactive routing protocols use a Least Overhead Routing Approach (LORA) [1][3] wherein an *s-d* route is discovered through a global broadcast flooding-based route discovery process and the discovered route is used as long as it exists. With node mobility, an *s-d* route determined to be optimal at a particular time instant need not remain optimal in the subsequent time instants, even though the route may continue to exist. Thus, with LORA, it is possible that the routing protocols continue to send data packets through sub-optimal routes. On the other hand, with ORA, even though, we could send data packets at the best possible route at any time instant, the cost of periodically discovering such a route may be significantly high. In dynamically changing network topologies, reactive on-demand routing protocols have been preferred over proactive protocols with respect to the routing control overhead incurred [4][5].

From another perspective, among the routing algorithms and protocols proposed for MANETs, routing based on a connected dominating set (CDS) has been recognized as a suitable approach in adapting quickly to the unpredictable fast-changing topology and dynamic nature of a MANET [6]. It is considered adaptable because as long as topological changes do not affect the structure of the CDS, there is no need to reconfigure the CDS since the routing paths based on the CDS would still be valid. A MANET is often represented as a unit disk graph [7] built of vertices and edges, where vertices signify nodes and edges signify bi-directional links that exist between any two nodes if they are within each other's transmission range. In a given graph representing a MANET, a CDS is a dominating set within the graph whose induced sub graph is connected. A dominating set of a graph is a vertex subset, such that every vertex is either in the subset or adjacent to a vertex in the subset [8]. Routing based on a CDS within a MANET means that routing control messages will be exchanged only amongst the CDS nodes and not broadcast by all the nodes in the network; this will reduce the number of unnecessary transmissions in routing [9].

There are multiple ways to form a CDS within a given MANET, and the algorithm used for CDS formation will affect the performance and lifetime of the CDS and the performance of the MANET as a whole. A popular approach in CDS formation is attempting to form the smallest possible CDS within a MANET, referred to as a minimum connected dominating set (MCDS). Reducing the size of the CDS will mean reducing the number of unnecessary transmissions. Unfortunately, the problem of determining a MCDS in an undirected graph like that of the unit disk graph is NP-complete [9][12]. Efficient heuristics [10][11][12] have been proposed to approximate the MCDS in wireless ad hoc networks. A common thread among these heuristics is to give the preference of CDS inclusion to nodes that have high neighborhood density. The MaxD-CDS heuristic [9] that we study in this paper is one such heuristic.

The objective of this paper is to study the impact of adopting the ORA and LORA strategies on minimum hop routes and the node size of the MCDS in MANETs. Minimum hop routing is a very widely adopted route selection principle of MANET routing protocols, belonging to both proactive and reactive categories. Likewise, the primary objective of a majority of the MCDS-based heuristics is to minimize the number of nodes constituting the CDS. As ORA determines the best optimal route at any time instant, our primary hypothesis is that the hop count of minimum hop routes and the node size of MCDS discovered under the LORA strategy would be greater than those discovered under the ORA strategy. Our secondary hypothesis is that the impact of ORA vs. LORA also depends on how long the communication structure is being used. We determine the percentage difference in the hop count of minimum hop *s-d* paths and the node size of the MCDS determined under the two strategies. We conduct extensive simulations under three different network densities and three different mobility models with three different levels of node mobility. The three mobility models [13] used are the Random Waypoint model, City Section model and Manhattan model. Even though performance comparison studies of individual proactive vs. reactive routing protocols as well as the different CDS algorithms are available in the literature, an extensive simulation based analysis on the impact of the ORA and LORA strategies on the minimum hop count of routes and the node size of the MCDS algorithms has not been conducted in the literature and therein lies our contribution through this paper.

The rest of the paper is organized as follows: Section 2 discusses the algorithms employed for determining minimum hop routes under the ORA and LORA strategies and also illustrates an example highlighting the difference between the two strategies and their impact on the hop count of *s-d* paths. Section 3 discusses the algorithms employed for determining MCDS under the ORA and LORA strategies and also illustrates an example highlighting the difference between the two strategies and their impact on the node size of the MCDS. Section 4 reviews the three different mobility models used in the simulations. Section 5 describes the simulation environment and presents the simulation results for hop count per *s-d* path, node size per MCDS, path lifetime and network connectivity. Section 6 concludes the paper and lists future work. Throughout the paper, the terms 'node' and 'vertex', 'edge' and 'link', 'path' and 'route' are used interchangeably. They mean the same.

## 2. DETERMINATION OF MINIMUM HOP ROUTES UNDER THE ORA AND LORA STRATEGIES

We use the notion of a mobile graph [14] defined as the sequence $G_M = G_1G_2 \dots G_T$ of static graphs that represent the network topology changes over the time scale $T$, representing the simulation time. We sample the network topology periodically, for every 0.25 seconds, which in reality could be the instants of data packet origination at the source. Each of the static graphs is a unit disk graph [7] of nodes and edges, wherein there exists an edge if and only if the Euclidean distance between the two constituent end nodes of the edge is within the transmission range of the nodes. We assume every node operates at a fixed transmission range, $R$.

For the ORA strategy, we determine the sequence of minimum hop *s-d* paths between a source node *s* and a destination node *d* by running the Breadth First Search (BFS) algorithm [15], starting from the source node *s*, on each of the static graphs of the mobile graph generated over the entire time period of the simulation. In the case of LORA, if we do not know a path from source *s* to destination *d* in static graph $G_i$, we run BFS (pseudo code in Figure 1), starting from node *s*, on $G_i$ and determine the minimum hop path $P_{s-d}$ from *s* to *d*. For subsequent static graphs $G_{i+1}$, $G_{i+2}$, ..., we simply test the presence of path $P_{s-d}$. We validate the existence of a path $P_{s-d}$ in static graph $G_j$ by testing the existence of every constituent edge of $P_{s-d}$ in $G_j$. If every constituent edge of $P_{s-d}$ exists in $G_j$, then the path $P_{s-d}$ exists in $G_j$. Otherwise, we run BFS on $G_j$, starting from the source node *s*, and determine a new *s-d* path $P_{s-d}$. This procedure is repeated until the end of the simulation time. The pseudo code of our algorithms to determine the minimum hop paths under the ORA and LORA strategies is given in Figures 2 and 3 respectively.

---

**Input:** Static Graph $G = (V, E)$, source node *s*, destination node *d*
**Auxiliary Variables/Initialization:** *Nodes-Explored* = Φ, *FIFO-Queue* = Φ
$\qquad\qquad\qquad\qquad\qquad$ ∀ node $v \in V$, *Parent* (*v*) = NULL
**Begin** Algorithm *BFS* (*G*, *s*, *d*)
$\quad$ *Nodes-Explored = Nodes-Explored* U {*s*}
$\quad$ *FIFO-Queue = FIFO-Queue* U {*s*}
$\quad$ **while** ( |*FIFO-Queue*| > 0 ) **do**
$\quad\quad$ node *u* = Dequeue(*FIFO-Queue*) // extract the first node
$\quad\quad$ **for** (every edge (*u*, *v*) ) **do** // i.e. every neighbor *v* of node *u*
$\quad\quad\quad$ **if** ( *v* ∉ *Nodes-Explored*) **then**
$\quad\quad\quad\quad$ *Nodes-Explored = Nodes-Explored* U {*v*}
$\quad\quad\quad\quad$ *FIFO-Queue = FIFO-Queue* U {*v*}
$\quad\quad\quad\quad$ *Parent* (*v*) = *u*
$\quad\quad\quad$ **end if**
$\quad\quad$ **end for**

**end while**
**if** ( | *Nodes-Explored* | = | *V* | ) **then**
    Path $P_{d\text{-}s}$ = {$d$}
    *temp-node* = $d$
    **while** (*Parent* (*temp-node*) != NULL) **do**
        $P_{d\text{-}s} = P_{d\text{-}s}$ U {*Parent* (*temp-node*)}
        *temp-node* = *Parent* (*temp-node*)
    **end while**
    Path $P_{s\text{-}d}$ = reverse($P_{d\text{-}s}$)
    **return** $P_{s\text{-}d}$
**end if**
**else**
    **return** NULL // no *s-d* path
**end if**
**End** Algorithm BFS

---

**Figure 1:** Breadth First Search (BFS) Algorithm to Determine Minimum Hop *s-d* Path

---

**Input:** $G_M = G_1 G_2 \dots G_T$, source *s*, destination *d*
**Auxiliary Variable:** *i*, Path $P_{s\text{-}d}$
**Initialization:** *i*=1; $P_{s\text{-}d}$ = NULL
**Begin** *ORA-MinHopPaths*
    **while** ($i \leq T$) **do**
        Path $P_{s\text{-}d}$ = BFS($G_i$, *s*, *d*)
        *i* = *i* + 1
    **end while**
**End** *ORA-MinHopPaths*

---

**Figure 2:** Pseudo Code to Find a Sequence of Minimum Hop *s-d* Paths under the ORA Strategy

---

**Input:** $G_M = G_1 G_2 \dots G_T$, source *s*, destination *d*
**Auxiliary Variables:** *i*, *j*, Path $P_{s\text{-}d}$
**Initialization:** *i*=1; *j*=1; $P_{s\text{-}d}$ = NULL
**Begin** *LORA-MinHopPaths*
    **while** ($i \leq T$) **do**

    **if** ($P_{s\text{-}d}$ != NULL) **then**
        **for** every edge (*u*, *v*) in $P_{s\text{-}d}$ **do**
            **if** ( (*u*, *v*) does not exist in $G_i$) **then**
                $P_{s\text{-}d}$ = NULL
            **end if**
        **end for**
    **end if**
    **if** ($P_{s\text{-}d}$ = NULL) **then**
        Path $P_{s\text{-}d}$ = BFS($G_i$, *s*, *d*)
    **end if**
    *i* = *i* + 1
    **end while**
**End** *LORA-MinHopPaths*

---

**Figure 3:** Pseudo Code to Find Sequence of Minimum Hop *s-d* Paths under the LORA Strategy

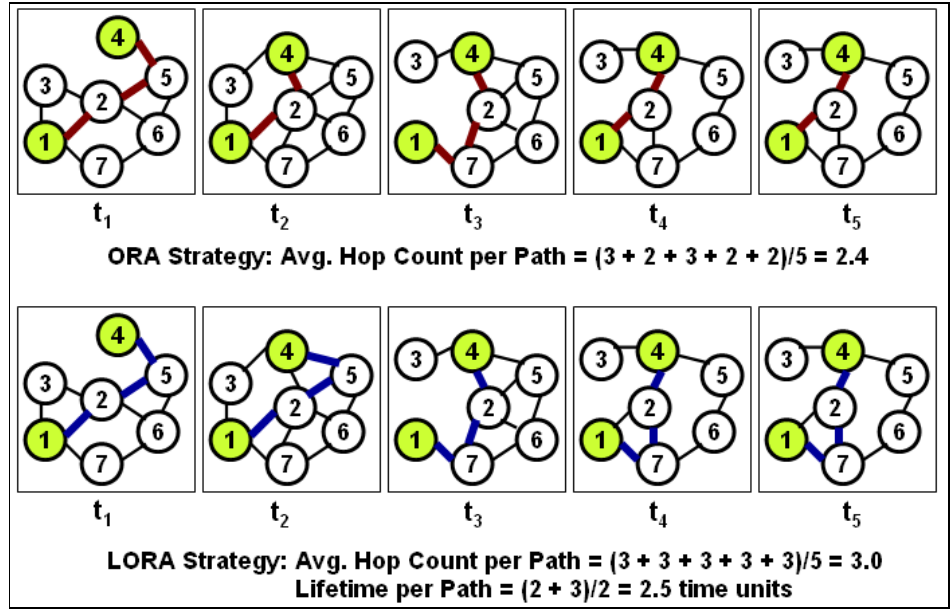

**Figure 4:** Example to Illustrate the ORA and LORA Strategies for Minimum Hop Routing

Figure 4 is an example to illustrate the difference between the ORA and LORA strategies with respect to minimum hop routing. We sample the network topology for five consecutive instants of time as shown. The source and destination node IDs are 1 and 4 respectively. We notice that under the LORA strategy, we could use path {1 – 2 – 4 – 5} for time instants $t_1$ and $t_2$ and path {1 – 7 – 2 – 4} for time instants $t_3$, $t_4$ and $t_5$ respectively. The paths {1 – 2 – 4 – 5} and {1 – 7 – 2 – 4} appear to be the best possible minimum hop paths at the time of discovery, i.e., at time instants $t_1$ and $t_3$ respectively. Nevertheless, after each of these paths is chosen at a particular time instant, we notice the emergence of relatively shorter paths (i.e., with a lower hop count) in the static graphs captured at subsequent time instants. But it is not possible to use these paths under the LORA strategy. With ORA, the strategy is to capture the minimum hop paths at every time instant.

## 3. DETERMINATION OF MINIMUM CONNECTED DOMINATING SETS UNDER THE ORA AND LORA STRATEGIES

The algorithm used to approximate a MCDS is referred to as the MaxD-CDS algorithm [9] as it prefers to include nodes that have a larger number of uncovered neighbors (density) to be part of the CDS. The MaxD-CDS algorithm uses the following principal data structures:

  (i)  *CDS-Node-List* – includes all nodes that are members of the CDS
  (ii) *Covered-Nodes-List* – includes all nodes that are in the CDS-Node-List and all nodes that are adjacent to at least one member of the CDS-Node-List.

Before we run the MaxD-CDS algorithm, we make sure the underlying network graph is connected by running the Breadth First Search (BFS) algorithm [15]; because, if the underlying network graph is not connected, we would not be able to find a CDS that will cover all the nodes in the network. We run BFS, starting with an arbitrarily chosen node in the network graph. If we are able to visit all the vertices in the graph, then the corresponding network is said to be connected. If the graph is not connected, we simply collect a snapshot of the network topology at the next time instant and start with the BFS test. The pseudo code for the BFS algorithm is given in Figure 5.

---

**Input:** Graph $G = (V, E)$
**Auxiliary Variables/Initialization:** *Nodes-Explored* = Φ, *FIFO-Queue* = Φ
**Begin** Algorithm *BFS* (*G, s*)
    *root-node* = randomly chosen vertex in *V*
    *Nodes-Explored* = *Nodes-Explored* U {*root-node*}
    *FIFO-Queue* = *FIFO-Queue* U {*root-node*}
    **while** ( |*FIFO-Queue*| > 0 ) **do**
        front-node *u* = Dequeue(*FIFO-Queue*) // extract the first node
        **for** (every edge (*u, v*) ) **do** // i.e. every neighbor *v* of node *u*
            **if** ( *v* ∉ *Nodes-Explored*) **then**
                *Nodes-Explored* = *Nodes-Explored* U {*v*}
                *FIFO-Queue* = *FIFO-Queue* U {*v*}
                *Parent* (*v*) = *u*
            **end if**
        **end for**
    **end while**

    **if** ( | *Nodes-Explored* | = | *V* | ) **then return** Connected Graph - true
    **else return** Connected Graph - false
    **end if**
**End** Algorithm BFS

---

**Figure 5:** Modified BFS Algorithm to Test for Graph Connectivity

The MaxD-CDS algorithm (pseudo code in Figure 6) outputs a *CDS-Node-List* based on a given input MANET graph. The first node to be included in the *CDS-Node-List* is the node with the maximum number of uncovered neighbors (any ties are broken arbitrarily). A CDS member is considered to be "covered", so a CDS member is additionally added to the *Covered-Nodes-List* as it is added to the *CDS-Node-List*. All nodes that are adjacent to a CDS member are also said to be covered, so the uncovered neighbors of a CDS member are also added to the *Covered-Nodes-List* as the member is added to the *CDS-Node-List*. To determine the next node to be added to the *CDS-Node-List*, we must select the node with the largest density amongst the nodes that meet the criteria for inclusion into the CDS. The criteria for CDS membership selection are the following: the node cannot already be a part of the CDS (*CDS-Node-List*), the node must be in the *Covered-Nodes-List*, and the node must have at least one uncovered neighbor (at least one neighbor that is not in the *Covered-Nodes-List*). Amongst the nodes that meet these criteria for CDS membership inclusion, we select the node with the largest density (i.e., the largest number of uncovered neighbors) to be the next member of the CDS. Ties are broken arbitrarily. This process is repeated until all nodes in the network are included in the *Covered-Nodes-List*. Once all nodes in the network are considered to be "covered", the CDS has been formed and the algorithm returns a list of the members included in the resultant MaxD-CDS (nodes in the *CDS-Node-List*).

---

**Input:** Graph $G = (V, E)$; *V* – vertex set, *E* – edge set
        Source vertex, *s* – vertex with the largest number of uncovered neighbors in *V*
**Auxiliary Variables and Functions:** *CDS-Node-List, Covered-Nodes-List, Neighbors*(*v*) for every *v* in *V*
**Output:** *CDS-Node-List*
**Initialization:** *Covered-Nodes-List* = {*s*}, *CDS-Node-List* = Φ
**Begin** Construction of *MaxD-CDS* (*G, s*)
    **while** ( |*Covered-Nodes-List*| < |*V*| ) **do**

Select a vertex $r \in$ *Covered-Nodes-List* and $r \notin$ *CDS-Node-List* such that $r$ has the largest number of uncovered neighbors that are not in *Covered-Nodes-List*

    *CDS-Node-List = CDS-Node-List* U {$r$}

      **for** all $u \in$ *Neighbors*($r$) and $u \notin$ *Covered-Nodes-List*
        *Covered-Nodes-List = Covered-Nodes-List* U {$u$}
      **end for**
    **end while**
  **return** *CDS-Node-List*
**End** Construction of *MaxD-CDS*

**Figure 6:** Pseudo Code for the Algorithm to Construct Maximum Density (MaxD)-based CDS

For the ORA strategy, we determine the sequence of MCDS by running the MaxD-CDS algorithm, starting from the source node $s$ – the node with the largest number of neighbors, on each of the static graphs of the mobile graph generated over the entire time period of the simulation. In the case of LORA, if we do not know a MCDS in static graph $G_i$, we run the MaxD-CDS algorithm, starting from the source node $s$ – the node with the largest number of neighbors in $G_i$ and determine the MCDS. For subsequent static graphs $G_{i+1}$, $G_{i+2}$, …, we simply test the presence of the MCDS. We validate a MCDS in a static graph $G_j$ by first testing the connectivity among the nodes that constitute the MCDS and then testing whether each non-MCDS node in $G_j$ is a neighbor of at least one node in the MCDS. If both these tests return true, then we consider the MCDS to exist in $G_j$. Otherwise, we run the MaxD-CDS algorithm on $G_j$, starting from a source node $s$ – the node with the largest number of neighbors in $G_j$ and determine a new MCDS. This procedure is repeated until the end of the simulation time. A pseudo code for the algorithm to validate a MCDS is given in Figure 7. The pseudo code of our algorithms to determine the MCDS under the ORA and LORA strategies is given in Figures 8 and 9 respectively.

---

**Input:** *CDS-Node-List* // Set of vertices part of the CDS
**Auxiliary Variables and Functions:**
  *CDS-Edge-List* – Set of edges, $\subseteq E$, between the vertices that are part of *CDS-Node-List*
  *connectedCDS* – Boolean variable that stores information whether *CDS-Node-List* and
               *CDS-Edge-List* form a connected sub graph of $G$.
**Output:** *true* or *false*
    *// true*, if the nodes in *CDS-Node-List* form a connected sub graph of $G$ and every vertex
      $v \notin$ *CDS-Node-List* is a neighbor of a vertex $u \in$ *CDS-Node-List*
    *// false*, if the nodes in *CDS-Node-List* do not form a connected sub graph of $G$ and/or
      there exists at least one vertex $v \notin$ *CDS-Node-List* that has no neighbor in *CDS-Node-List*
**Initialization:** *CDS-Edge-List* = Φ
**Begin** CDS-Validation (*CDS-Node-List*, time instant $t$)
  **for** every pair of vertices $u$, $v \in$ *CDS-Node-List* **do**
    **if** there exists an edge ($u$, $v$)$\in E$ at time instant $t$ **then**
      *CDS-Edge-List = CDS-Edge-List* U {($u$, $v$)}
    **end if**
  **end for**
  *connectedCDS* = Breadth-First-Search(*CDS-Node-List*, *CDS-Edge-List*)
  **if** *connectedCDS* = true **then**

    **for** every vertex $v \notin$ *CDS-Node-List* **do**
      **if** there exists no edge ($u$, $v$)$\in E$ where $u \in$ *CDS-Node-List* at time instant $t$ **then**

       **return** *false*
    **end if**
  **end for**

    **return** *true*
  **end if**

  **return** *false* // if *connectedCDS* = false

**End** CDS-Validation

---

**Figure 7:** Pseudo Code for the CDS Validation Algorithm

---

**Input:** $G_M = G_1 G_2 \ldots G_T$
**Auxiliary Variable:** $i$, $MCDS_i$
**Initialization:** $i$=1; $MCDS_i$ = NULL

**Begin** *ORA-MCDS*
    **while** ($i \leq T$) **do**
      Choose the source node $s$ – the node with the largest number of neighbors in $G_i$
      **if** *BFS* ($G_i$, $s$) returns **true**
        $MCDS_i$ = MaxD-CDS($G_i$, $s$)
      **end if**
      $i = i + 1$
    **end while**
**End** *ORA-MCDS*

---

**Figure 8:** Pseudo Code to Determine a Sequence of MCDS under the ORA Strategy

---

**Input:** $G_M = G_1 G_2 \ldots G_T$, source $s$, destination $d$
**Auxiliary Variables:** $i$, $j$, $MCDS$
**Initialization:** $i$=1; $j$=1; $MCDS$ = NULL
**Begin** *LORA-MCDS*
    **while** ($i \leq T$) **do**
      **if** ($MCDS$ != NULL) **then**
          **if** ( CDS-Validation (*CDS-Node-List* of *MCDS*, time instant $i$) returns **false**) **then**
            $MCDS$ = NULL
          **end if**
      **end if**
      **if** ($MCDS$ = NULL) **then**
        Choose the source node $s$ – the node with the largest number of neighbors in $G_i$
        **if** *BFS* ($G_i$, $s$) returns **true**
          $MCDS$ = MaxD-CDS($G_i$, $s$)
        **end if**
      **end if**

      $i = i + 1$

    **end while**
**End** *LORA-MCDS*

---

**Figure 9:** Pseudo Code to Determine Minimum Hop *s-d* Paths under the LORA Strategy

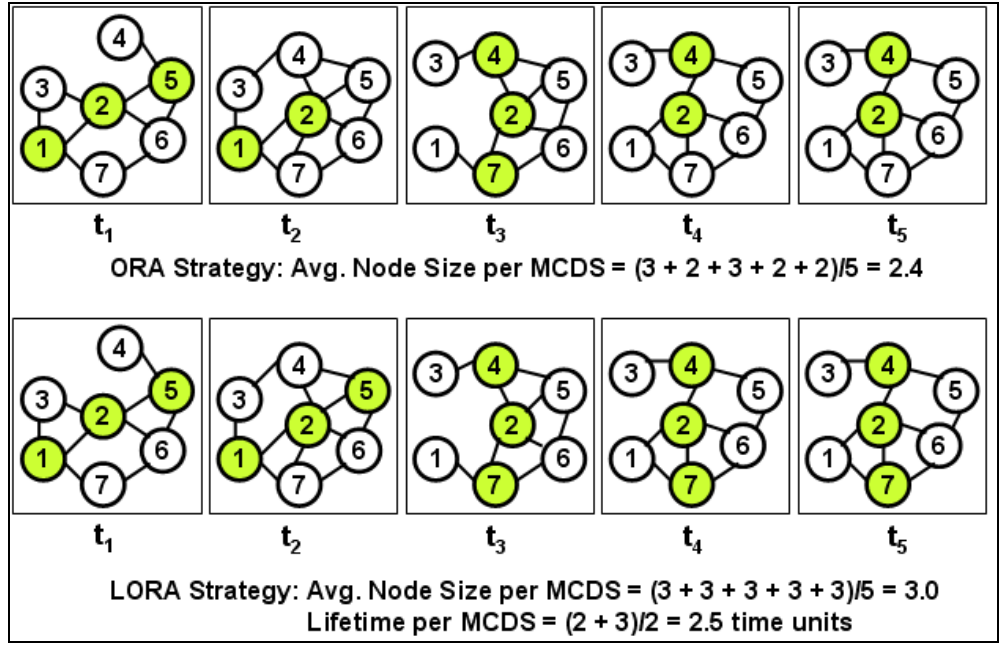

**Figure 10:** Example to Illustrate the ORA and LORA Strategies for Determining MCDS

Figure 10 is an example to illustrate the difference between the ORA and LORA strategies with respect to determining MCDS. We sample the network topology for five consecutive instants of time as shown. To determine a MCDS on a particular network topology, we start with the vertex (node ID 2 in all the cases) that has the largest number of neighbors. While deciding whether a covered node can be part of the *CDS-Node-List* of the MCDS, we include the covered node with the largest number of uncovered neighbors. Any tie in this case is broken in favor of the covered node that has the lowest ID. We notice that under the LORA strategy, we could use the MCDS comprising of nodes {2, 1, 5} with edges {1 − 2, 2 − 5} for time instants $t_1$ and $t_2$ and the MCDS comprising of nodes {2, 4, 7} with edges {2 − 4, 2 − 7} for time instants $t_3$, $t_4$ and $t_5$ respectively. The average MCDS node size under the LORA approach is 3.0 as there are three nodes in the MCDS used in each of the five time instants. On the other hand, under the ORA strategy, we determine MCDS comprising of nodes {2, 1, 5}, {2, 1}, {2, 4, 7}, {2, 4}, {2, 4} at time instants $t_1$, $t_2$, $t_3$, $t_4$ and $t_5$ respectively. The average MCDS node size is 2.4. Notice that the absence of link 2 − 1 in the graph at time instant $t_3$ forced us to choose the MCDS with nodes {2, 4, 7} at $t_3$; once this link appears at time instants $t_4$ and $t_5$, by adopting the ORA strategy – we could reduce the number of nodes in the MCDS from three to two; whereas, by adopting the LORA strategy, we end up continuing to stay with a MCDS comprising of three nodes. Updating the MCDS for every time instant helps to reduce the number of constituent CDS nodes; however, with a significant control overhead. Using a CDS with a larger number of constituent nodes leads to redundant retransmissions in the case of flooding using the CDS. This illustrates the difference and trade off between the ORA and LORA strategies.

## 4. REVIEW OF MOBILITY MODELS

All the three mobility models assume the network is confined within fixed boundary conditions. The Random Waypoint mobility model assumes that the nodes can move anywhere within a network region. The City Section and the Manhattan mobility models assume the network to be divided into grids: square blocks of identical block length. The network is thus basically composed of a number of horizontal and vertical streets. Each street has two lanes, one for each direction (north and south direction for vertical streets, east and west direction for horizontal streets). A node is allowed to move only along the grids of horizontal and vertical streets.

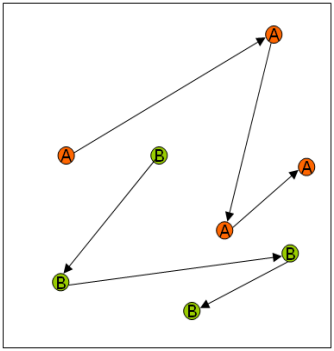

**Figure 11:** Movement under
Random Waypoint Model

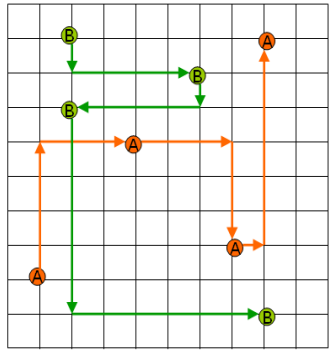

**Figure 12:** Movement under
City Section Model

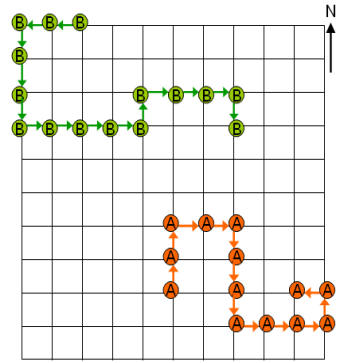

**Figure 13:** Movement under
Manhattan Mobility Model

### 4.1. Random Waypoint Mobility Model

Initially, the nodes are assumed to be placed at random locations in the network. The movement of each node is independent of the other nodes in the network. The mobility of a particular node is described as follows: The node chooses a random target location to move. The velocity with which the node moves to this chosen location is uniformly and randomly selected from the interval $[v_{min},...,v_{max}]$. The node moves in a straight line (in a particular direction) to the chosen location with the chosen velocity. After reaching the target location, the node may stop there for a certain time called the pause time. The node then continues to choose another target location and moves to that location with a new velocity chosen again from the interval $[v_{min},...,v_{max}]$. The selection of each target location and a velocity to move to that location is independent of the current node location and the velocity with which the node reached that location. In Figure 11, we observe that nodes *A* and *B* move independent of each other, in random directions with randomly chosen velocities.

### 4.2. City Section Mobility Model

Initially, the nodes are assumed to be randomly placed in the street intersections. Each street (i.e., one side of a square block) is assumed to have a particular speed limit. Based on this speed limit and the block length, one can determine the time it would take move in the street. Each node placed at a particular street intersection chooses a random target street intersection to move. The node then moves to the chosen street intersection on a path that will incur the least amount of travel time. If two or more paths incur the least amount of travel time, the tie is broken arbitrarily. After reaching the targeted street intersection, the node may stay there for a pause time and then again choose a random target street intersection to move. The node then moves towards the new chosen street intersection on the path that will incur the least amount of travel time. This procedure is repeated independently by each node. In Figure 12, the movement of two nodes *A* and *B* according to the City Section mobility model has been illustrated.

### 4.3. Manhattan Mobility Model

Initially, the nodes are assumed to be randomly placed in the street intersections. The movement of a node is decided one street at a time. To start with, each node has equal chance (i.e., probability) of choosing any of the streets leading from its initial location. In Figure 13, to start with, node *A* has 25% chance to move in each of the four possible directions (east, west, north or south), where as node *B* can move only either to the west, east or south with a 1/3 chance for each direction. After a node begins to move in the chosen direction and reaches the next street intersection, the subsequent street in which the node will move is chosen probabilistically. If a node can continue to move in the same direction or can also change directions, then the node has 50% chance of continuing in the same direction, 25% chance of turning to the east/north and 25% chance of turning to the west/south, depending on the direction of the previous movement. If a node has only two options, then the node has an equal (50%) chance of exploring either of the two options. For example, in Figure 13, once node *A* reaches the rightmost boundary of the network, the node can either move to the north or to the south, each with a probability of 0.5 and the node chooses the north direction. After moving to the street intersection in the north, node *A* can either continue to move northwards or turn left and move eastwards, each with a probability of 0.5. If a node has only one option to move (this occurs when the node reaches any of the four corners of the network), then the node has no other choice except to explore that option. For example, in Figure 13, we observe node *B* that was traveling westward, reaches the street intersection, which is the corner of the network. The only option for node B is then to turn to the left and proceed southwards.

## 5. SIMULATIONS

Simulations have been conducted in a discrete-event simulator implemented by the author in Java. Network dimensions are 1000m x 1000m. For the Random Waypoint mobility model, we assume the nodes can move anywhere within the network. For the City Section and Manhattan mobility models, we assume the network is divided into grids: square blocks of length (side) 100m. The network is thus basically composed of a number of horizontal and vertical streets. Each street has two lanes, one for each direction (north and south direction for vertical streets, east and west direction for horizontal streets). A node is allowed to move only along the grids of horizontal and vertical streets. The wireless transmission range of a node is 250m. The network density is varied by performing the simulations with 50 (low density), 100 (moderate density) and 150 (high density) nodes. The node velocity values used for each of the three mobility models are 2.5 m/s (about 5 miles per hour), 12.5 m/s (about 30 miles per hour) and 25 m/s (about 60 miles per hour), representing scenarios of low, moderate and high node mobility respectively. For the Random Waypoint mobility model, we assume $v_{min} = v_{max}$.

We obtain a centralized view of the network topology by generating mobility trace files for 1000 seconds under each of the three mobility models. The network topology is sampled for every 0.25 seconds to generate the static graphs and the mobile graph. Two nodes *a* and *b* are assumed to have a bi-directional link at time *t*, if the Euclidean distance between them at time *t* (derived using the locations of the nodes from the mobility trace file) is less than or equal to the wireless transmission range of the nodes. Each data point in Figures 14 through 19 and in Tables 1 to 6 is an average computed over 5 mobility trace files and 20 randomly selected *s-d* pairs from each of the mobility trace files. The starting time of each *s-d* session is uniformly distributed between 1 to 20 seconds.

The following performance metrics are evaluated:

- *Percentage Network Connectivity*: The percentage network connectivity indicates the probability of finding an *s-d* path between any source *s* and destination *d* in networks for a given density and a mobility model. Measured over all the *s-d* sessions of a simulation run, this metric is the ratio of the number of static graphs in which there is an *s-d* path to the total number of static graphs in the mobile graph.
- *Average Route Lifetime*: The average route lifetime is the average of the lifetime of all the static paths of an *s-d* session, averaged over all the *s-d* sessions.
- *Average Hop Count*: The average hop count is the time averaged hop count of a mobile path for an *s-d* session, averaged over all the *s-d* sessions. The time averaged hop count for an *s-* *d* session is measured as the sum of the products of the number of hops per static *s-d* path and the lifetime of the static *s-d* path divided by the number of static graphs in which there existed a static *s-d* path. For example, if a mobile path spanning over 10 static graphs comprises of a 2-hop static path $p_1$, a 3-hop static path $p_2$, and a 2-hop static path $p_3$, with each existing for 2, 3 and 5 seconds respectively, then the time-averaged hop count of the mobile path would be (2*2 + 3*3 + 2*5) / 10 = 2.3.

- *CDS Node Size*: This is a time-averaged value of the number of nodes included in the sequence of minimum connected dominating sets used over the entire duration of the simulation.

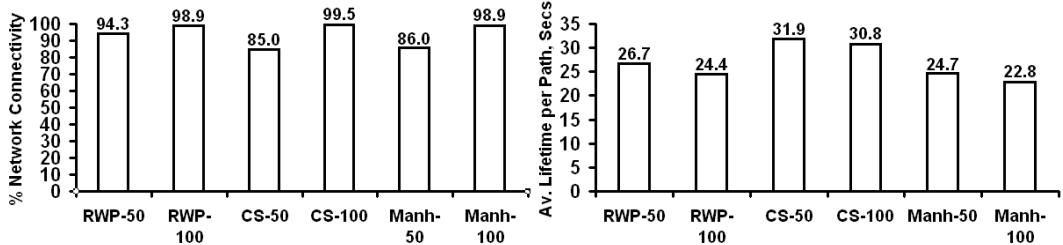

**Figure 14:** % Connectivity (vel = 2.5 m/s)    **Figure 15:** Lifetime per *s-d* Path (vel = 2.5 m/s)

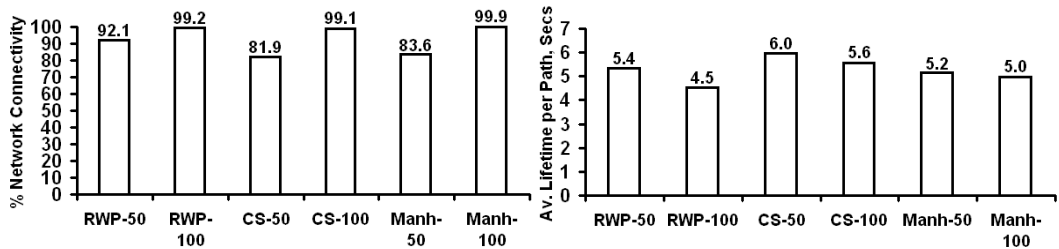

**Figure 16:** % Connectivity (vel = 12.5 m/s)    **Figure 17:** Lifetime per *s-d* Path (vel = 12.5 m/s)

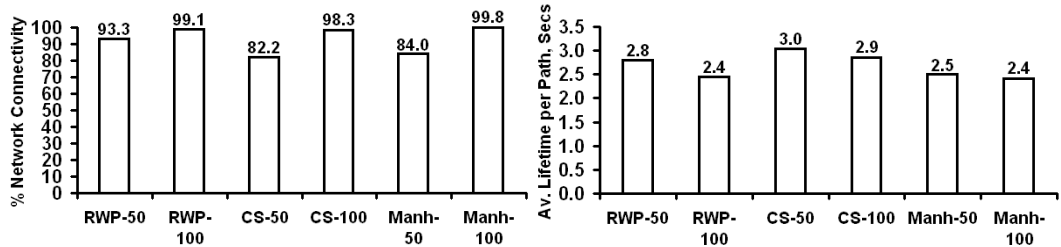

**Figure 18:** % Connectivity (vel = 25 m/s)    **Figure 19:** Lifetime per *s-d* Path (vel = 25 m/s)

## 5.1. Network Connectivity

The percentage network connectivity (refer Figures 14, 16 and 18) is not dependent on the routing strategy (ORA or LORA) and is dependent only on the mobility model, the level of node mobility and network density. It is quite natural to observe that for a given mobility model and level of node mobility, the percentage network connectivity increases with increase in network density. In low density networks (50 nodes), the Random Waypoint model provided the largest network connectivity for a given level of node mobility; the City Section and Manhattan models yielded a relatively lower network connectivity, differing as large as by 11% . This can be attributed to the constrained motion of the nodes only along the streets of the network. On the other hand, as we increase the network density (100 node scenarios), the City Section model and/or the Manhattan model yielded network connectivity equal or larger than that incurred with the Random Waypoint model. As more nodes are added to the streets, the probability of finding source-destination routes at any point of time increases significantly. It is also interesting to observe that for a given network density, the network connectivity provided by each of the three mobility models almost remained the same for different values of node velocity. Hence, network connectivity is mainly influenced by the number of nodes in the network and their initial random distribution. The randomness associated with the mobility models ensure that node velocity is not a significant factor influencing network connectivity.

## 5.2. Route Lifetime

The average route lifetime (Figures 15, 17 and 19) is measured only for routes discovered under the LORA strategy as routes are determined for every static graph under the ORA strategy. With LORA, a route is used as long as it exists. The average route lifetime of minimum hop routes is mainly influenced by node velocity and to a lesser extent by the mobility model and network density, in this order. For a given node velocity and network density, minimum hop routes determined under the City Section model had the largest lifetime and those determined under the Manhattan model had the smallest lifetime except the scenario of 100 nodes with 12.5 m/s velocity, wherein the Random Waypoint model yielded routes with the lowest average lifetime. For a given node velocity, the difference in the average lifetime of routes between the City Section model and the other two mobility models increase with increase in network density. The City Section model yielded a route lifetime that is 8-20% and 17-26% more than that discovered under the Random Waypoint model in low and high density networks respectively. Compared to the Manhattan model, the City Section model yielded routes that have 15-30% and 12-35% larger lifetime in low and high density networks respectively. For a given mobility model, the route lifetime seem to decrease proportionately with increase in node velocity. As we increase the node velocity from 2.5 m/s to 25 m/s, the average lifetime of minimum hop routes determined under a particular mobility model approximately reduced to $1/10^{th}$ of their value at low node velocity.

## 5.3. Hop Count of Minimum Hop Routes

For each mobility model, node velocity and network density, we observe that minimum hop routes discovered under the LORA strategy has a larger hop count than those discovered under the ORA strategy. But, the increase in the hop count is not substantial and is within 12%. This indicates that if the on-demand MANET routing protocols based on the LORA strategy are designed meticulously with minimum hop routing as the primary routing principle, they could discover routes that have at most 12% larger hop count than those discovered by the ORA-based proactive routing protocols. Among the three mobility models, the maximum increase in the hop count under the LORA strategy vis-à-vis the ORA strategy is observed with the Random Waypoint model and the lowest increase in the hop count is observed with the Manhattan model. However, with regards to the absolute values of the hop count, the minimum hop routes determined under the Random Waypoint model have the smallest hop count and those determined under the Manhattan model have the largest hop count.

For a given mobility model, the hop count of the minimum hop routes determined for a particular network density does not seem to be much influenced with different levels of node mobility. For a given mobility model and node velocity, we also observe that under both the ORA and LORA strategies, the average hop count of minimum hop routes decreases with increase in network density. This can be attributed to the reasoning that with a larger number of nodes in the network, there is a larger probability of finding an *s-d* path involving only fewer nodes that lie on the path from the source to the destination. The decrease in the hop count of minimum hop routes with increase in network density is very much appreciable for the Manhattan model compared to the other two mobility models.

Another interesting observation is that for a given network density, the percentage increase in the average hop count per minimum hop *s-d* path decreases with increase in node mobility. This can be attributed to significant decrease in the lifetime of the *s-d* routes with increase in node mobility. At higher node mobility, the sub-optimal routes do not exist for a longer time and the sequence of routes determined under the LORA strategy starts getting closer to the sequence of routes determined under the ORA strategy. This effect is more predominant in the case of MCDS as the lifetime per CDS under the LORA strategy is significantly smaller than the lifetime per *s-d* path.

**Table 1:** Average Hop Count per *s-d* Path under Random Waypoint Mobility Model

| Node Velocity | 50 Node Network | | | 100 Node Network | | | 150 Node Network | | |
|---|---|---|---|---|---|---|---|---|---|
| | ORA | LORA | Percent Increase | ORA | LORA | Percent Increase | ORA | LORA | Percent Increase |
| 2.5 m/s | 2.36 | 2.63 | 11.51% | 2.27 | 2.50 | 10.40% | 2.21 | 2.43 | 9.95% |
| 12.5 m/s | 2.40 | 2.65 | 10.14% | 2.36 | 2.58 | 9.46% | 2.25 | 2.46 | 9.33% |
| 25 m/s | 2.40 | 2.63 | 9.44% | 2.31 | 2.52 | 9.31% | 2.24 | 2.44 | 8.93% |

**Table 2:** Average Hop Count per *s-d* Path under City Section Mobility Model

| Node Velocity | 50 Node Network | | | 100 Node Network | | | 150 Node Network | | |
|---|---|---|---|---|---|---|---|---|---|
| | ORA | LORA | Percent Increase | ORA | LORA | Percent Increase | ORA | LORA | Percent Increase |
| 2.5 m/s | 2.66 | 2.86 | 7.51% | 2.45 | 2.71 | 10.40% | 2.24 | 2.50 | 11.60% |
| 12.5 m/s | 2.85 | 3.07 | 7.66% | 2.70 | 2.93 | 8.68% | 2.55 | 2.80 | 9.80% |
| 25 m/s | 2.83 | 3.04 | 7.39% | 2.60 | 2.82 | 8.47% | 2.37 | 2.60 | 9.70% |

**Table 3:** Average Hop Count per *s-d* Path under Manhattan Mobility Model

| Node Velocity | 50 Node Network | | | 100 Node Network | | | 150 Node Network | | |
|---|---|---|---|---|---|---|---|---|---|
| | ORA | LORA | Percent Increase | ORA | LORA | Percent Increase | ORA | LORA | Percent Increase |
| 2.5 m/s | 3.31 | 3.60 | 8.81% | 3.08 | 3.34 | 8.38% | 2.75 | 3.03 | 10.12% |
| 12.5 m/s | 3.37 | 3.60 | 6.90% | 3.00 | 3.26 | 8.60% | 2.67 | 2.93 | 9.82% |
| 25 m/s | 3.51 | 3.74 | 6.60% | 3.03 | 3.27 | 7.94% | 2.52 | 2.76 | 9.35% |

## 5.4. Node Size per Minimum Connected Dominating Sete

For each mobility model, the average node size per MCDS determined under the LORA strategy is slightly higher than that determined under the ORA strategy. But, the increase is very minimal and is only within 6%. This implies that the number of retransmissions incurred by adopting the sequence of MCDS determined under the LORA strategy will not be substantially higher than those incurred using the sequence of MCDS determined under the ORA strategy. On the other hand, there would be a significant control overhead in updating the MCDS for every time instant. Hence, the LORA strategy could always be the preferred strategy to determine and use MCDS in MANETs.

With respect to the absolute magnitude of the MCDS Node Size under the three mobility models, we observe that the MCDS Node Size determined under the Random Waypoint model is always the smallest and the MCDS Node Size determined under the Manhattan mobility model is always the largest under the different conditions of node mobility and network density. The MCDS Node Size determined under the City Section mobility model is 16%, 20%-25% and 22%-24% larger than that determined under the Random Waypoint model in conditions of low, moderate and high network density respectively. The MCDS Node Size determined under the Manhattan mobility model is 26%-36%, 31%-34% and 30%-34% larger than that determined under the Random Waypoint model in conditions of low, moderate and high network density respectively.

**Table 4:** Average Node Size per MCDS under Random Waypoint Mobility Model

| Node Velocity | 50 Node Network | | | 100 Node Network | | | 150 Node Network | | |
|---|---|---|---|---|---|---|---|---|---|
| | ORA | LORA | Percent Increase | ORA | LORA | Percent Increase | ORA | LORA | Percent Increase |
| 2.5 m/s | 9.80 | 10.12 | 3.27% | 10.17 | 10.62 | 4.47% | 10.09 | 10.65 | 5.55% |
| 12.5 m/s | 9.88 | 10.23 | 3.52% | 9.93 | 10.24 | 3.12% | 10.41 | 10.75 | 3.25% |
| 25 m/s | 9.54 | 9.94 | 4.19% | 9.81 | 10.15 | 3.47% | 10.21 | 10.51 | 2.93% |

**Table 5:** Average Node Size per MCDS under City Section Mobility Model

| Node Velocity | 50 Node Network | | | 100 Node Network | | | 150 Node Network | | |
|---|---|---|---|---|---|---|---|---|---|
| | ORA | LORA | Percent Increase | ORA | LORA | Percent Increase | ORA | LORA | Percent Increase |
| 2.5 m/s | 11.43 | 11.78 | 3.06% | 12.18 | 12.62 | 3.61% | 12.58 | 13.07 | 3.89% |
| 12.5 m/s | 11.37 | 11.79 | 3.69% | 12.18 | 12.65 | 3.86% | 12.91 | 13.31 | 3.09% |
| 25 m/s | 11.15 | 11.47 | 2.87% | 12.22 | 12.46 | 1.96% | 12.68 | 12.91 | 1.81% |

**Table 6:** Average Node Size per MCDS under Manhattan Mobility Model

| Node Velocity | 50 Node Network | | | 100 Node Network | | | 150 Node Network | | |
|---|---|---|---|---|---|---|---|---|---|
| | ORA | LORA | Percent Increase | ORA | LORA | Percent Increase | ORA | LORA | Percent Increase |
| 2.5 m/s | 12.42 | 12.89 | 3.78% | 13.33 | 13.61 | 2.10% | 13.53 | 13.95 | 3.10% |
| 12.5 m/s | 12.90 | 13.42 | 4.03% | 13.34 | 13.76 | 3.15% | 13.58 | 13.98 | 2.94% |
| 25 m/s | 13.01 | 13.32 | 2.38% | 13.11 | 13.39 | 2.14% | 13.45 | 13.73 | 2.08% |

As observed in the case of minimum hop routes, for a given network density, the percentage increase in the MCDS Node Size decreases with increase in node mobility. This can be attributed to the decrease in the MCDS lifetime by factors of 4 to 5 and 8 to 9 with increase in the node velocity from 2.5 m/s to 12.5 m/s and 25 m/s respectively. For a given condition of node mobility and network density, the lifetime per MCDS is only $1/3^{rd}$ to $1/4^{th}$ of the lifetime per *s-d* path determined under similar conditions. Hence, compared to the minimum hop routes, the *CDS-Node-List* of the sequence of MCDS formed under the LORA strategy fast coincides with that of the sequence of MCDS formed under the ORA strategy.

## 6. CONCLUSIONS AND FUTURE WORK

Our hypothesis that there would be difference in the hop count of minimum hop routes and the node size of the minimum connected dominating sets (MCDS) discovered under the ORA and LORA strategies has been observed to be true through extensive simulations, the results of which are summarized in Tables 1 through 6. However, the difference is not significantly high and is within 6-12% for minimum hop routes and at most 6% for MCDS, depending mainly on the mobility model employed and the level of node mobility and to a lesser extent on the network density. With respect to absolute values, the Random Waypoint model yields minimum hop routes with the smallest hop count and MCDS with the smallest node size; whereas, the Manhattan model yields minimum hop routes with the largest hop count and MCDS with the largest node size. With respect to the increase in the hop count of minimum hop routes due to the use of LORA strategy vis-à-vis the ORA strategy, we observe that the Random Waypoint model incurs the maximum increase and the Manhattan model incurs the smallest increase. The City Section model is ranked in between the two mobility models with regards to the absolute value of the hop count and the relative increase in the hop count with the LORA strategy. In the context of the MCDS, the percentage increase in the number of nodes per MCDS due to the use of LORA vis-à-vis ORA is about the same for all the three mobility models. With regards to the route lifetime, the minimum hop routes determined under the City Section model are relatively more stable (i.e. have larger lifetime) compared to the other two mobility models.

Another interesting observation is that for a given network density, the percentage increase in the average hop count per minimum *s-d* path and the number of nodes per MCDS decreases with increase in node mobility. This can be attributed to the significant decrease in the lifetime of the *s-d* routes and the MCDS with increase in node mobility. This effect is more predominant in the case of MCDS as the lifetime per MCDS under the LORA strategy is significantly smaller than the lifetime per *s-d* path. Hence, compared to the minimum hop routes, the *CDS-Node-List* of the sequence of MCDS formed under the LORA strategy fast coincides with that of the sequence of MCDS formed under the ORA strategy. As future work, we will be extending this study and will examine the impact of the ORA vs. LORA strategies and the three mobility models on minimum-hop based multicast routing, minimum-link based multicast Steiner trees, as well as node-disjoint and link-disjoint multi-path routing for MANETs.

**Author**

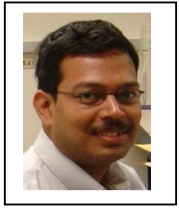

Dr. Natarajan Meghanathan is an Assistant Professor of Computer Science at Jackson State University. He graduated with MS and PhD degrees in Computer Science from Auburn University and The University of Texas at Dallas respectively. He has authored more than 100 peer-reviewed publications. He has received federal grants from the U. S. National Science Foundation the Army Research Lab. He is serving in the editorial board of several international journals and in the organization/program committees of several international conferences. His research interests are: Ad hoc Networks, Sensor Networks, Network Security, Graph Theory and Bioinformatics.